# Effect of the sEMG electrode (re)placement and feature set size on the hand movement recognition

Nadica Miljković[a,*], Milica S. Isaković[1a,b,c]

[a]University of Belgrade - School of Electrical Engineering, Bulevar kralja Aleksandra 73, 11000 Belgrade, Serbia

[b]Innovation Center, University of Belgrade – School of Electrical Engineering, Bulevar kralja Aleksandra 73, 11000 Belgrade, Serbia

[c] Tecnalia Serbia Ltd., Deligradska 9/39, 11000 Belgrade, Serbia

**DATA AVAILABILITY:** Recorded data are freely available on Zenodo repository: "Surface electromyogram (sEMG) dataset recorded from forearm for 9 hand movements and three electrode array positions", doi: 10.5281/zenodo.4039550 under CC Attribution 4.0 International licence.

**ABSTRACT**

Repositioning of recording electrode array across repeated electromyography measurements may result in a displacement error in hand movement classification systems.

In order to examine if the classifier re-training could reach satisfactory results when electrode array is translated along or rotated around subject's forearm for varying number of features, we recorded surface electromyography signals in 10 healthy volunteers for three types of grasp and 6 wrist movements. For feature extraction we applied principal component analysis and the feature set size varied from one to 8 principal components. We compared results of re-trained classifier with results from leave-one-out cross-validation classification procedure for three classifiers: LDA (Linear Discriminant Analysis), QDA (Quadratic Discriminant Analysis), and ANN (Artificial Neural Network).

Our results showed that there was no significant difference in classification accuracy when the array electrode was repositioned indicating successful classification re-training and optimal feature set selection. The results also indicate expectedly that the number of principal components plays a key role for acceptable classification accuracy ~90%. For the largest dataset (9 hand movements), LDA and QDA outperformed ANN, while for three grasping movements ANN showed promising results. Interestingly, we showed that interaction between electrode array position and the feature set size is not statistically significant.

This study emphasizes the importance of testing the interaction of factors that influence classification accuracy and classifier selection altogether with their impact independently in order to establish guiding principles for design of hand movement recognition system.

Data recorded for this study are stored on Zenodo repository (doi: 10.5281/zenodo.4039550).

---

[1] Milica S. Isaković was affiliated with the University of Belgrade – School of Electrical Engineering when the work was done. This is not her current address.

# 1 INTRODUCTION[2]

Hand movement recognition systems based on application of surface electromyography (sEMG) are predominately affected and limited by: (1) acquisition setup, (2) protocol design, (3) signal pre-processing, (4) feature extraction, and (5) classifier design, and to date researchers have offered numerous advances in order to enhance system functionality [1-10].

Isaković et al. [5] showed that significant enhancement can be obtained by increasing the number of principal components (PCs) in systems where Principal Component Analysis (PCA) is used for feature extraction and applied on benchmark sEMG data recorded in intact subjects and obtained from the publicly available NinaPro database [11-12]. These results suggest that instead of using two PCs for dimension reduction as in [11] for obtaining considerably low error rates, only one PC more (three PCs in total) can significantly increase classification accuracy in three sets of movements (>10%). This compelling finding was further tested on sEMG data recorded in three intact subjects when electrode array was relocated in proximal direction [13] and results showed that when using three PCs electrode dislocation did not affect significantly classification accuracy (<5%). To the best of our knowledge, the optimal number of PCs and its exact relation to the sEMG electrode array dislocation to operation of hand movement recognition systems is still unknown. Hence, exploring this relation could possibly improve the knowledge on the selection of feature set size, its relationship to common problem of electrode dislocation, and on possibility to compensate for electrode dislocation.

The main idea of this paper is to test whether reasonable classification accuracy (~90%) can be reached by re-training the classifier when electrode array is dislocated in relation to the feature set size. Although, NinaPro database [11-12] provides perfectly usable benchmark data, it does not incorporate the measurements with electrode dislocation, and therefore we could not use existing database. We recorded sEMG signals for hand movements previously presented in [5, 11, 13] for three positions of sEMG electrode array with 8 channels arranged circumferentially. Three positions of electrode array were selected according to the previously reported findings of common electrode array displacement in neuroprosthesis [14]. Then, the accuracy for three classifiers was tested for all three positions, for three sets of hand movements (three grasps, 6 wrist movements, and for joined grasps and wrist movements – 9 hand movements in total), and for 1-8 PCs.

The results presented here can advance areas related to the surface electromyography (sEMG) acquisition setup that includes electrode array (re)positioning and offer optimal number of features, all being required for successful sEMG-based hand movement classification that can be used in human machine interfaces (HMI). We had no intention to limit the recommendations to neuroprosthesis control solely. However, we present related work mostly in the area of sEMG-based neuroprosthesis control as one of the most important and oldest (> 60 years) hand movement recognition systems [15].

The main limitation of the state-of-the-art hand movement recognition systems for neuroprosthesis is the limited number of discernible movements that leads to reduced operational intuitiveness in spite of captive improvements [4, 6-7, 10]. Here, we aimed at addressing two challenges and their relation in a controlled laboratory setting: (1) sEMG electrode repositioning and (2) selection of adequate number of PCs.

## 1.1 sEMG acquisition setup challenge: Electrode repositioning

Day-to-day positioning of the sEMG array electrode can produce undesirable electrode displacements. It is shown that accuracy changes due to the electrode displacement and can be affected by: inter-electrode spacing, the number of electrodes, and the number of classified motions [16]. Matching sEMG records in consecutive user utilizations requires re-placing the electrode array over identical recording locations. Such protocol is demanding and involves specialized knowledge of muscle anatomy. Therefore, accommodation of training and/or classification strategy to sEMG electrodes repositioning has been proposed [17]. There is a need for a wide variety of examples (that result from electrode displacement) to be presented to a classifier (during a training phase) in order to apply sEMG-based pattern recognition with satisfactory performance [16]. Offered solutions include enlargement of the training set as proposed in [16-17], daily calibration aiming at accommodation to displacement changes as introduced in [8], and transfer learning approach has also been proposed [18, 19] to compensate for electrode dislocation.

---

[2] **Non-standard abbreviations** include types of hand postures: relaxation (R), spherical power grasp (PS), three finger sphere grasp (3F), two finger prismatic grasp (PP), wrist flexion (FL), wrist extension (EX), radial deviation (RD), ulnar deviation (UD), forearm rotation i.e., pronation (PR), and supination (SU).

In order to enhance more knowledge of electrode dislocation on classifier performance, we tested the effect of feature set size on dislocation of sEMG array electrode that was translated along and rotated around subject's forearm when classifier is re-trained as we did not aim at pre-training of classifier as proposed recently [18].

**1.2 Feature extraction challenge: Optimal number of principal components**

PCA is an orthogonal data transformation method commonly used for feature extraction from multi-channel sEMG signals [4-5, 9, 11, 13, 20] in order to reduce the dimensionality of multidimensional data and to reveal data patterns. Another PCA application is for multi-channel sEMG-based assessment of muscle sub-modules termed neuromuscular compartments [21], but this application is out of our scope. We applied PCA for feature extraction as we assumed that the temporal-spatial information contained within muscle crosstalk on the forearm might implicitly add class discriminatory information as suggested in [4-5, 11]. Isaković et al.[5] found that the proportion of variance present in the first three PCs compared to first two PCs extracted from the sEMG data from the NinaPro database provides a statistically significant increase in classification accuracy which is promising. However, the optimal number of PCs was not evaluated and here we present, in detail, evaluation of an optimal number of PCs in relation to sEMG array electrode repositioning.

**1.3 Study objectives**

With the aim to evaluate the effects of electrode array dislocation on PCA-based classification feature selection and classification accuracy of three classifiers: (1) Linear Discriminant Analysis (LDA), Quadratic Discriminant Analysis (QDA), and Artificial Neural Network (ANN), we measured 8-channel sEMG from 10 volunteers for 9 hand movements in a controlled laboratory setting with similar measurement paradigm as in [2]. Recorded data provided a deeper look at the exact influence of electrode array re-positioning (translational and rotational) on the robustness of feature extraction for three hand movement sets.

**2 METHODS AND MATERIALS**

All processing steps were performed in Matlab ver. 2013b (Mathworks Inc., Natick, USA). Data are available on general purpose open-access repository Zenodo [22].

**2.1 Intact subjects**

We measured sEMG signals from 10 healthy volunteers with no known neuromuscular or skeletal disorders. All volunteers signed an Informed Consent and the study was performed in compliance with the Code of Ethics of the University of Belgrade, which provides guidelines for studies involving human beings and is in accordance with the Declaration of Helsinki. The demographic data with anthropometric measurements (forearm length and circumference) for all subjects are presented in Table 1.

Table 1, Demographic data with anthropometric measurements for 10 healthy volunteers (ID1-ID10). The abbreviations are M - Male, F - Female, L - left, and R - right.

| Subject ID | Height [cm] | Weight [kg] | Sex [M/F] | Age [years] | Forearm length [cm] | Forearm circumference [cm] | Dominant arm [L/R] |
| --- | --- | --- | --- | --- | --- | --- | --- |
| ID1 | 168 | 58 | F | 23 | 24 | 20.0 | L |
| ID2 | 175 | 65 | F | 28 | 25 | 24.0 | R |
| ID3 | 193 | 85 | M | 27 | 29 | 26.0 | R |
| ID4 | 180 | 68 | M | 26 | 28 | 27.0 | R |
| ID5 | 168 | 53 | F | 25 | 24 | 23.0 | R |
| ID6 | 186 | 83 | M | 23 | 27 | 26.5 | R |
| ID7 | 164 | 50 | F | 22 | 23 | 22.0 | R |
| ID8 | 178 | 67 | M | 22 | 27 | 24.5 | R |
| ID9 | 175 | 66 | M | 27 | 28 | 24.0 | L |
| ID10 | 165 | 51 | F | 25 | 24 | 19.0 | R |
| All | 175.2±9.4 | 64.6±12.2 | 5M+5F | 24.8±2.2 | 25.9±2.1 | 23.6±2.7 | 8R+2L |

## 2.2 sEMG measurement setup

An array of 8 Ag/AgCl electrode pairs arranged uniformly and circumferentially around the forearm of the dominant arm was used for bipolar sEMG measurement.

It is assumed that 8 channels would be sufficient for adequate feature extraction and classification accuracy, especially as our electrode array was placed more distally than in [11] in order to cover the muscle bulge – more similar to the [4]. Our approach presents a compromise between array location presented in [11] and [4], while having in mind the SENIAM recommendation to place sEMG electrodes over "the most prominent bulge of the muscle belly" [23] and the fact that muscle cross-talk would occur.

We used disposable, pre-gelled, self-adhesive, and with active circular area sEMG Skintact F-TC1 (Leonhard Lang GmbH, Innsbruck, Austria) electrodes (2 x 8). The position of the electrode array was chosen according to the proximate bulks of *extensor carpi radialis* and *flexor carpi ulnaris* muscles which correspond to the one-third of the distance between the elbow and wrist [4]. Prior to the electrode placement, the skin was cleaned with the Nuprep abrasive gel (Bio-Medical Instruments, Inc., Warren, MI, USA). A Skintact F-TC1 reference electrode was placed over the *articulatio cubiti* bone.

The signals were digitized with the USB AD converter NI6212 (National Instruments, Inc., Austin, USA) with 16 bits resolution and sampling rate of 1000 samples per second. The signals were amplified (Biovision, Inc., Wehrheim, Germany) with gain of 1000. A recent study [24] compared acquisition setups and concluded that no significant difference was observed between low- and high-priced systems. We used the Biovision sEMG system in the mid-price range. For signal acquisition, we adopted a custom program previously created in the LabVIEW (National Instruments, Inc., Austin, USA) environment [25]. Subjects were prompted to follow the software feedback instructions with LED and audio cues. Each movement repetition lasted for 5 s with LED indicator set to the "ON" state and it was followed by 3 s long rest position with LED set to the "OFF" state.

## 2.3 Measurement protocol

Prior to each measurement session, an instruction video was presented on a computer screen. The instruction video consisted of a movement being performed (explanatory movement videos are available at https://www.youtube.com/playlist?list=PLI3SYeiSufnBo6UDAZt9NJO9ecb-InJqb). The subjects were seated on a chair with adjustable heights, and instructed to regulate the chair height so that the angle between forearm and upper arm was approximately 90 degrees.

The subjects were asked to perform the following movements from the reference resting position termed relaxation, R: (1) spherical power grasp, PS, (2) three finger sphere grasp, 3F, (3) two finger prismatic grasp, PP, (4) wrist flexion, FL, (5) wrist extension, EX, (6) radial deviation, RD, (7) ulnar deviation, UD, and then forearm rotation i.e. (8) pronation, PR, and (9) supination, SU, [11, 26]. For the PP, we provided a plastic cylinder (height of 9 cm with a diameter of 3 cm). The PS and 3F movements were performed with a ball (weight of 95 g with diameter of 6 cm). An image of a subject's position during relaxation (R) and grasping movements is presented in Fig. 1.

We recorded 10 trials for each of the 9 tasks for the following three positions of the electrode array presented in Fig. 2:

    Position #1: "standard" position where the electrode array was placed on the forearm longitudinally in the *proximal/distal* direction; for details see 2.2 sEMG measurement setup,

    Position #2: 1 cm translation (3.9±0.3% of the forearm length in all subjects) of the electrode array from Position #1 in the *proximal* direction, and

    Position #3: 2 cm rotation (8.6±1.0% of the forearm circumference in all subjects) of the electrode array from Position #1 in the *medial* direction.

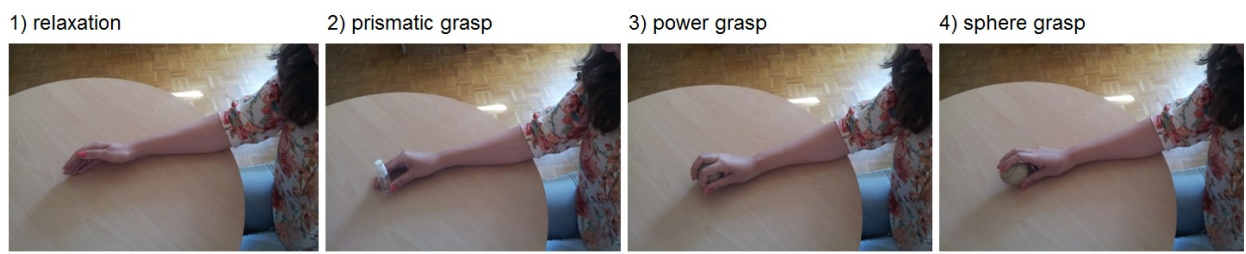

Figure 1, Photo impression of subject's posture: 1) relaxation (R); and for the three grasp modes: 2) two finger prismatic grasp (PP), 3) spherical power grasp (PS), and 4) three finger sphere grasp (3F).

The shifts of 1 cm in the *proximal* and of 2 cm in the *medial* direction were chosen based on the findings from [14] that (1) 2 cm shift represents one of the worst possible conditions when using a socket sEMG electrode array and that (2) 1 cm shift or less is likely caused by the fitting of sEMG electrode array.

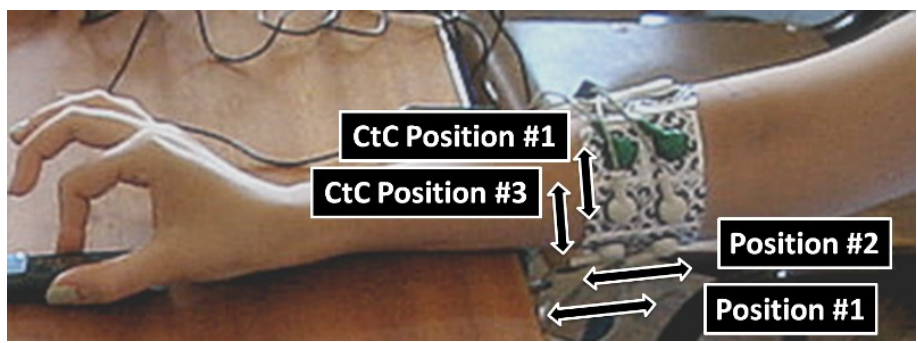

Figure 2, Photo impression of the "standard" Position #1 and two dislocations: (1) Position #2 is Position #1 translated 1 cm in the *proximal* direction and (2) Position #3 is Position #1 rotated 2 cm in the *medial* direction. Center to center (CtC) locations correspond to the distances of active circular areas of sEMG electrodes from neighboring sEMG channels.

The distance between two electrodes in the *medial/lateral* direction is constrained by the design and by the forearm circumference, so we kept the electrodes equally spaced in the *medial/lateral* direction and at fixed distance of 2 cm in the *proximal/distal* direction. For all Positions #1-3, sEMG electrodes were taken off and re-placed. All signals for each subject were recorded in a single recording session. In order to avoid fatigue, frequent 5-20 min breaks were provided. The total recording duration for obtaining Information Consent, sEMG array electrode positioning, movement execution, and repositioning was approximately 2 hours.

**2.4 sEMG data processing and feature extraction**

The sEMG signals were filtered with a notch filter (50 Hz) in order to eliminate power line interference, followed by the first-order modified differential infinite impulse response (IIR) filter in order to remove the baseline offset. The sEMG signals were then rectified and filtered with $3^{rd}$ order low-pass Butterworth filter with cut-off frequency of 5 Hz in order to generate sEMG envelopes. Subsequently, we applied segmentation and averaging as described elsewhere [5, 11] in order to obtain a single feature per movement. Each repetition of the movement in time domain (termed posture) and the following relaxation (termed pause) was divided into three segments of equal duration, in order to retain the most representative time frame from the central segment. With assumption that signal is stationary on extracted segments, we averaged samples across the central segment in order to obtain single feature per movement which resulted in 20 features (10 per posture and 10 per pause) for each subject/movement/electrode array position combination (20 x 10 x 9 x 3 = 5400 features in total). The preprocessing procedure is illustrated in Fig. 3, comprising rectified signal, sEMG envelope, and features of pause and posture for a single movement repetition. The features for individual subjects were subsequently normalized in order to obtain a zero mean and unit standard deviation. Next, PCA was applied to the normalized features.

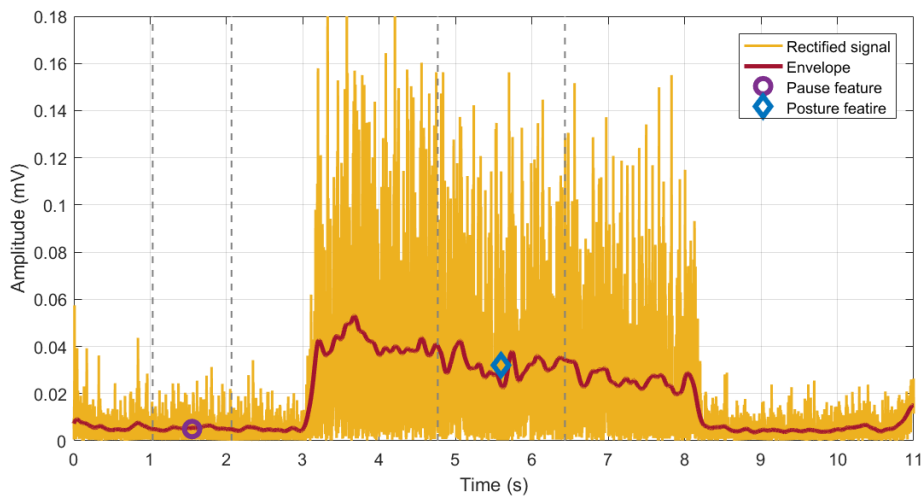

Figure 3, Example of a single movements repetition (subject ID1, wrist flexion recorded in Position #1, channel 7) illustrating applied pre-processing procedure. Raw sEMG signal was filtered to remove power line interference and baseline offset, and rectified (yellow signal). The envelope (red signal) was obtained by low-pass filtering. Samples from central segments of both pause and posture were averaged to obtain a single feature (pause – purple circle, posture – blue diamond).

**2.5 Classification, statistical analysis, and optimal number of PCs**

We used LDA (Linear Disriminant Analysis), QDA (Quadratic Discriminant Analysis), and ANN (Artificial Neural Network) for classification of extracted features.

We used QDA as it provided satisfactory results in previous studies [5, 8]. QDA allows non-linear i.e., quadratic decision boundaries between data. However, as this classifier showed significantly worse performance than LDA and ANN in both able-bodied subjects and patients in [1], we decided to include both LDA and ANN. LDA and QDA are classifiers with linear and non-linear discriminant functions and they are very simple to compute as the rule of classification is to find the class which maximizes the discriminant function. LDA is surely most commonly used classifier, although reported LDA performance for hand movement classification varied from the worst to the best, probably depending on other factors [1, 27, 28]. It has been suggested that ANN outperformed other classifiers for myoelectric-based control in [28]. ANN as discriminative classifier commonly has higher specificity, but requires computational complexity and can be subject to over-fitting on relatively small samples of data [29]. ANN was formed as a two-layer feed-forward network with sigmoid activation function in its one hidden layer comprising 8 neurons and softmax activation function in an output layer, and trained with scaled conjugate gradient backpropagation. We applied two approaches: (1) where subject independency was not ensured (as it was assumed that only one subject would use the system) and (2) we added evaluation framework with leave-one-out cross-validation in order to highlight user specific sensitivity. In the first approach, the feature dataset was split into two sets (training 80% and test 20%) for LDA and QDA, and three sets (training 80%, validation 10% ,and test 10%) for ANN classifier. The second approach relied on the procedure that was repeated for each subject: the classifier was trained in 9 subjects and tested in the remaining subject.

The accuracy of the classifier was compared for 8 different feature sets (1-8 PCs). Simultaneously, we examined the classification accuracy for three movement sets: (1) three grasps, (2) 6 wrist movements, and (3) all movements (i.e.,9 hand movements = 3 grasps + 6 wrist movements). Muscle relaxation was used as an additional input for each movement set. For all combinations of features (1-8 PCs) and three movement sets, we performed classification for Positions #1-3. We re-trained the classifier for Positions #1-3 in order to select the optimal feature set size for different electrode array positions i.e., we enabled classifier to re-run and to be trained from the beginning (during this process, the features and the classifier type remained the same, only dataset changed) in order to correct for the classifier accuracy degradation [30]. The outcome measure of the classification performance was classification accuracy i.e., the percentage of correctly classified repetitions of the movements for each of the 10 subjects, three sets of movement, 8 sets of features, and three electrode array positions.

All statistical tests were performed individually for each set of movements. Two-way analysis of variance (ANOVA) and Tukey's honestly significant difference criterion for multiple comparison tests were performed with a variable number of features (1-8 PCs) and for all electrode array positions (Positions #1-3) as factors. In order to determine the optimal number of features, statistically significant differences in mean classification accuracy among 8 sets of features for each electrode array position were tested using one-way ANOVA and Tukey's honestly significant difference criterion for post-hoc pairwise comparisons.. We introduced criteria for the optimal number of PCs when the classification accuracy could not be significantly changed by providing additional PC and this was not related to the highest classification accuracy. An equivalent analysis was performed to explore the differences between the electrode array positions for each number of PCs used as classification features. The threshold for the statistical significance was set at $p < 0.05$.

We compared performance of LDA, QDA, and ANN in order to choose the most appropriate classifier. Statistical comparison was carried out for each electrode array position (P1, P2, and P3) and classification method (classifier re-trained for each individual subject and leave-one-out cross-validation procedure) with 2-way ANOVA with two factors: (1) classifier type (LDA, QDA, and ANN) and (2) number of PCs (1-8). For the chosen classifier, we presented confusion matrices and classifier sensitivity and specificity for each movement.

# 3 RESULTS

In Tables 2-4, classification accuracies with standard deviations for three classifiers and two classification methods (re-trained classifier for each individual subject and leave-one-out cross-validation) are presented for three grasping movements, 6 wrist movements, and 9 hand movements, respectively.

Table 2, Classification accuracy (mean ± standard deviation for 10 subjects) for 8 sets of features comprising 1-8 PCs, three positions of electrode array (P1, P2, P3), three types of classifiers (LDA, QDA, ANN) and two classification methods (Classifier re-trained for each individual subject, leave-one-out cross-validation) in case of set of three grasping movements. Bold values present optimal number of features for selected classifier.

|  |  | Classification accuracy in percents for re-trained classifier for individual subjects ||||||||
|  |  | P1 ||| P2 ||| P3 |||
|  |  | LDA | QDA | ANN | LDA | QDA | ANN | LDA | QDA | ANN |
| --- | --- | --- | --- | --- | --- | --- | --- | --- | --- | --- |
| Number of PCs | 1 | 48.8±22.4 | 51.3±19.0 | 63.5±14.1 | 45.0±22.2 | 51.3±18.1 | 65.0±9.0 | 47.5±14.2 | 43.8±20.6 | 67.0±13.3 |
|  | 2 | 67.5±32.4 | 68.8±24.5 | 79.3±14.7 | 61.3±19.9 | 62.5±24.3 | 81.8±14.0 | 62.5±25.7 | 57.5±23.0 | 77.0±10.8 |
|  | 3 | **85.0±16.5** | **83.8±14.5** | **90.5±10.6** | 82.5±22.2 | 78.8±18.7 | **88.0±15.7** | 71.3±18.7 | 70.0±18.8 | 86.5±13.3 |
|  | 4 | 92.5±8.7 | 87.5±14.4 | 94.3±11.5 | 91.3±8.4 | **83.8±15.6** | 95.8±6.4 | **81.3±14.7** | **80.0±15.8** | **96.0±4.6** |
|  | 5 | 95.0±8.7 | 96.3±8.4 | 96.8±8.6 | 87.5±11.8 | 88.8±12.4 | 97.8±4.0 | 91.3±10.3 | 91.3±11.9 | 98.0±2.3 |
|  | 6 | 100.0±0.0 | 98.8±4.0 | 97.5±7.9 | 93.8±8.8 | 92.5±12.1 | 91.8±23.5 | 95.0±10.5 | 95.0±8.7 | 96.8±7.8 |
|  | 7 | 100.0±0.0 | 100.0±0.0 | 92.3±23.6 | 100.0±0.0 | 100.0±0.0 | 100.0±0.0 | 97.5±5.3 | 98.8±4.0 | 99.3±1.7 |
|  | 8 | 100.0±0.0 | 100.0±0.0 | 95.0±10.5 | 100.0±0.0 | 100.0±0.0 | 100.0±0.0 | 100.0±0.0 | 100.0±0.0 | 97.5±7.9 |
|  |  | Classification accuracy in percents for leave-one-out cross-validation ||||||||
|  |  | P1 ||| P2 ||| P3 |||
|  |  | LDA | QDA | ANN | LDA | QDA | ANN | LDA | QDA | ANN |
| Number of PCs | 1 | 55.3±7.3 | 52.3±6.7 | 55.2±2.1 | 53.8±13.0 | 53.3±14.3 | 52.9±1.5 | 50±15.3 | 51.5±14.1 | 52.6±1.8 |
|  | 2 | 56.0±10.2 | 52.7±9.4 | 55.4±6.5 | 58.3±14.2 | **60.0±12.2** | 62.3±5.3 | 51±11.6 | 47.8±10.0 | 57.5±6.1 |
|  | 3 | 56.0±8.4 | 53.3±10.4 | 60.8±8.0 | 60.8±14.8 | 64.5±11.8 | 67.9±7.1 | 63.3±12.0 | 61.3±9.9 | 66.4±10.4 |
|  | 4 | **63.3±11.4** | **62.3±15.1** | 67.6±8.3 | 61.8±15.1 | 63.5±13.2 | 70.5±2.2 | **68.3±12.8** | 68.3±12.1 | 69.9±9.1 |
|  | 5 | 68.5±9.6 | 66.5±11.6 | 66.2±11.9 | 66.0±15.1 | 67.0±18.0 | 73.0±6.3 | 79.3±13.6 | **78.3±13.2** | 79.6±7.5 |
|  | 6 | 71.5±11.5 | 71.8±12.1 | **78.9±3.4** | 73.5±13.6 | 74.0±13.8 | **78.0±10.8** | 81.5±10.9 | 81.0±9.9 | 84.0±10.1 |
|  | 7 | 71.3±11.9 | 72.0±10.6 | 76.0±7.4 | 72.8±11.8 | 76.0±13.8 | 79.8±7.2 | 83.5±10.9 | 82.0±10.3 | **90.3±1.7** |
|  | 8 | 76.5±13.1 | 75.3±11.9 | 79.9±6.4 | 78.5±10.9 | 79.5±15.4 | 81.4±13.0 | 84.5±10.0 | 84.5±9.8 | 84.5±8.6 |

The results of two-way ANOVA with classifier type and number of PCs as main effects were used to determine the most suitable classifier for the further analysis. . As expected, number of PCs has statistically significant effect on classification accuracy for all positions and classification approaches. However, classifier performance was affected by the choice of hand movement set in a following manner:
1. Three grasping movements – for individual approach (classifier re-trained for individual subjects) there is no difference for P1 among classifiers, and for P2 and P3 ANN performs better than LDA and QDA. For leave-one-out cross-validation ANN performs better than LDA and QDA.
2. Six wrist movements – for both classification methods and for three electrode positions there was no statistically significant difference among classifiers.

Nine hand movements –for individual approach there is no difference among classifiers for P1 and P2, and for P3 ANN performs significantly worse than LDA. For leave-one cross-validation in P1 and P2 LDA and QDA are significantly better than ANN, while in P3 only LDA is significantly better than ANN.
Since there was no consistency in classifier performance, we decided to continue further analysis with QDA, as it was employed in our previous study [5].

Table 3, Classification accuracy (mean ± standard deviation for 10 subjects) for 8 sets of features comprising 1-8 PCs, three positions of electrode array (P1, P2, P3), three types of classifiers (LDA, QDA, ANN) and two classification methods (Classifier re-trained for each individual subject, leave-one-out cross-validation) in case of set of 6 wrist movements. Bold values present optimal number of features for selected classifier.

| | | Classification accuracy in percents for re-trained classifier for individual subjects | | | | | | | | |
| --- | --- | --- | --- | --- | --- | --- | --- | --- | --- | --- |
| | | P1 | | | P2 | | | P3 | | |
| | | LDA | QDA | ANN | LDA | QDA | ANN | LDA | QDA | ANN |
| Number of PCs | 1 | 32.1±19.1 | 28.6±10.6 | 41.9±7.0 | 32.1±14.0 | 27.9±8.6 | 42.0±5.8 | 31.4±14.0 | 27.9±13.7 | 46.0±8.4 |
| | 2 | 44.3±14.2 | 45.7±17.9 | 53.4±17.1 | 57.1±15.4 | 47.9±16.5 | 58.7±13.9 | 46.4±12.7 | 47.9±11.7 | 53.9±9.8 |
| | 3 | 60.7±16.9 | 58.6±18.1 | 71.7±12.6 | 70.7±18.6 | 67.1±13.6 | 71.0±9.0 | 72.1±17.6 | 72.1±18.0 | 71.4±10.7 |
| | 4 | 77.9±17.6 | 80.0±10.0 | **82.0±14.0** | 80.0±13.0 | 79.3±9.8 | **82.6±13.7** | 85.0±11.4 | 80.0±17.4 | 80.7±11.3 |
| | 5 | **91.4±8.8** | **90.0±9.6** | 91.6±9.0 | **90.0±10.8** | **91.4±4.5** | 93.6±8.9 | 90.0±10.8 | **87.1±11.1** | **92.0±7.8** |
| | 6 | 97.1±5.0 | 95.7±6.0 | 92.4±14.5 | 93.6±8.6 | 97.1±5.0 | 98.1±4.5 | 95.7±7.7 | 92.1±11.4 | 97.0±6.6 |
| | 7 | 97.9±3.5 | 99.3±2.3 | 94.3±10.0 | 97.9±3.5 | 98.6±3.0 | 84.3±23.8 | 99.3±2.3 | 97.9±4.8 | 95.0±8.0 |
| | 8 | 99.3±2.3 | 100.0±0.0 | 89.4±23.6 | 99.3±2.3 | 99.3±2.3 | 91.4±15.4 | 98.6±4.5 | 99.3±2.3 | 95.6±6.8 |
| | | Classification accuracy in percents for leave-one-out cross-validation | | | | | | | | |
| | | P1 | | | P2 | | | P3 | | |
| | | LDA | QDA | ANN | LDA | QDA | ANN | LDA | QDA | ANN |
| Number of PCs | 1 | 25.7±6.7 | 26.9±5.4 | 32.5±1.6 | 33.3±5.9 | 32.3±6.5 | 35.8±0.8 | 32.7±7.0 | 30.3±3.7 | 32.6±1.3 |
| | 2 | 49.4±11.8 | 50.0±13.6 | 47.5±4.3 | 47.7±11.0 | 47.6±13.5 | 48.8±5.4 | 46.7±9.9 | 47.7±7.6 | 44.6±7.8 |
| | 3 | **62.6±16.2** | **62.7±17.1** | 56.0±8.3 | **58.4±14.3** | **56.4±14.2** | 53.3±8.8 | **56.9±10.9** | **56.7±8.5** | 52.6±7.2 |
| | 4 | 69.3±19.6 | 67.3±21.7 | 59.4±8.7 | 61.9±14.0 | 59.1±12.9 | 59.1±4.5 | 59.0±9.2 | 58.7±7.6 | 57.6±6.3 |
| | 5 | 72.9±19.6 | 73.7±19.1 | 64.6±12.7 | 65.0±12.0 | 63.0±11.2 | 53.7±15.5 | 62.3±10.5 | 61.9±9.7 | **62.1±7.7** |
| | 6 | 73.7±20.4 | 73.3±19.0 | 60.5±15.8 | 69.1±12.4 | 67.4±10.2 | 62.8±12.5 | 61.3±11.0 | 60.6±10.6 | 58.8±10.9 |
| | 7 | 74.7±15.9 | 75.9±16.9 | 67.8±17.0 | 70.3±12.1 | 69.6±9.6 | **69.1±7.1** | 60.7±10.2 | 58.9±12.9 | 56.3±7.5 |
| | 8 | 77.0±14.9 | 78.0±15.1 | **79.1±8.1** | 70.4±10.6 | 70.4±8.4 | 68.4±8.3 | 64.9±9.3 | 63.4±11.0 | 60.2±10.2 |

**3.1 Optimal number of PCs and results of statistical tests for three movement sets for re-training QDA**

For the set of three grasps, the optimal number of features (PCs) is three in case of Position #1 and average classification accuracy for all subjects is 83.8±14.5%, while it increases to four for both Position #2 and Position #3, with average accuracies of 83.8±15.7% and 80.0±15.8%, respectively. Two-way ANOVA revealed that feature set size (p < 0.01) and the sEMG electrode array location (p = 0.03) are statistically significant factors while their interaction was non significant (p = 0.97).

An optimal number of features for the set of 6 wrist movements is 5 for Position #1, Position #2, and Position #3 with average classification accuracies of 90.0±9.6%, 91.4±4.5% and 87.1±11.1%, respectively. For this set, two-way ANOVA revealed that feature set size is statistically significant (p < 0.01), while the sEMG electrode array position was not statistically significant (p = 0.75) so as the interaction of these two factors (p = 0.85).

For all 9 hand movements (6 wrist + three grasping movements), the optimal number of PCs in this case is 5 for Position #1 with classification accuracy of 85.5±11.4% and 6 for Position #2 and Position #3, with average classification accuracies of 89.5±8.3%, and 90.0±11.5%, respectively. Two-way ANOVA revealed that only feature set size was statistically significant (p < 0.01) while electrode array location (p = 0.24) and interaction of these two factors (p > 0.99) were non-significant.

The one-way ANOVA for all feature sets (1-8 PCs) showed no statistically significant difference in mean classification accuracy for three electrode positions #1-3 for all movement sets (9 hand, 6 wrists, and three grasping movements).

Table 4, Classification accuracy (mean ± standard deviation for 10 subjects) for 8 sets of features comprising 1-8 PCs, 3 positions of electrode array (P1, P2, P3), 3 types of classifiers (LDA, QDA, ANN) and 2 classification methods (Classifier re-trained for each individual subject, leave-one-out cross-validation) in case of set of 9 hand movements. Bold values present optimal number of features for selected classifier.

|  |  | Classification accuracy in percents for re-trained classifier for individual subjects | | | | | | | | |
|---|---|---|---|---|---|---|---|---|---|---|
|  |  | P1 | | | P2 | | | P3 | | |
|  |  | LDA | QDA | ANN | LDA | QDA | ANN | LDA | QDA | ANN |
| Number of PCs | 1 | 21.5±12.5 | 22.0±8.9 | 28.9±5.1 | 21.5±10.8 | 19.0±6.1 | 29.6±5.0 | 23.5±11.6 | 20.0±10.3 | 28.6±5.5 |
|  | 2 | 33.5±16.0 | 41.5±16.3 | 45.2±9.9 | 38.0±18.1 | 38.0±13.8 | 43.0±7.5 | 32.0±9.8 | 32.0±6.7 | 42.9±6.8 |
|  | 3 | 56.0±14.5 | 57.0±17.5 | 57.1±7.5 | 61.0±17.3 | 56.5±17.5 | 55.0±14.3 | 54.5±15.7 | 54.5±12.3 | 55.4±9.1 |
|  | 4 | 76.0±11.3 | 70.0±13.9 | 65.2±13.5 | 75.5±15.9 | 71.5±13.8 | 72.0±11.8 | 76.5±13.6 | 69.5±16.6 | 66.3±9.3 |
|  | 5 | **86.0±9.1** | 85.5±11.4 | 83.1±11.7 | 83.0±15.5 | 82.5±13.4 | **79.0±10.2** | 85.5±11.9 | 82.5±13.0 | 73.8±8.5 |
|  | 6 | 97.0±3.5 | 94.0±5.7 | 84.6±11.8 | 91.5±7.5 | **89.5±8.3** | 85.1±12.8 | 92.5±7.5 | **90.0±11.5** | 81.5±15.8 |
|  | 7 | 99.5±1.6 | 99.5±1.6 | 91.7±6.3 | 99.0±3.2 | 97.0±3.5 | 92.1±8.8 | 97.5±4.3 | 97.0±4.2 | 81.9±18.5 |
|  | 8 | 99.5±1.6 | 100.0±0.0 | 91.0±12.9 | 99.5±1.6 | 99.0±2.1 | 91.9±11.3 | 99.0±2.1 | 99.5±1.6 | 89.7±9.6 |
|  |  | Classification accuracy in percents for leave-one-out cross-validation | | | | | | | | |
|  |  | P1 | | | P2 | | | P3 | | |
|  |  | LDA | QDA | ANN | LDA | QDA | ANN | LDA | QDA | ANN |
| Number of PCs | 1 | 17.4±4.6 | 19.7±5.5 | 21.4±3.8 | 22.2±5.2 | 20.9±6.7 | 22.7±1.4 | 21.6±4.9 | 20.7±3.4 | 21.4±2.8 |
|  | 2 | 33.5±6.2 | 33.4±6.5 | 33.3±3.7 | 34.1±9.0 | 35.2±9.8 | 32.1±7.1 | 31.9±7.8 | 30.5±6.5 | 30.8±4.0 |
|  | 3 | 46.0±10.1 | 45.9±10.6 | 39.9±5.6 | 43.6±12.2 | 43.2±10.7 | 34.5±5.2 | 44.3±11.0 | 44.2±11.0 | 39.8±9.5 |
|  | 4 | **52.4±12.5** | **52.6±14.3** | 46.1±6.4 | 49.1±12.3 | 46.6±11.3 | 42.3±7.5 | **49.8±12.5** | **48.8±11.6** | 44.3±8.4 |
|  | 5 | 56.9±14.0 | 56.4±13.3 | 46.2±7.2 | 52.1±10.0 | **50.7±12.7** | 44.3±9.7 | 55.0±13.0 | 53.9±12.6 | **50.5±6.2** |
|  | 6 | 60.3±16.3 | 58.4±15.7 | **49.1±9.2** | **60.1±8.5** | 58.1±11.5 | 48.8±12.5 | 56.3±11.2 | 55.5±10.7 | 51.8±4.9 |
|  | 7 | 62.6±12.8 | 62.0±14.2 | 46.3±15.9 | 61.4±8.4 | 60.4±10.2 | **55.8±7.5** | 58.0±10.2 | 56.2±9.9 | 52.9±7.4 |
|  | 8 | 67.1±11.3 | 66.1±12.5 | 51.5±11.5 | 66.0±9.1 | 64.9±10.6 | 55.6±10.6 | 62.0±9.5 | 60.6±9.7 | 51.9±7.3 |

The highest classification accuracy of 90.0±9.6% among sets of hand movements and electrode array positions was obtained for 6 wrist movements for Position #1 and optimal number of 5 PC components. This is evidently higher from classification accuracy of three grasps of 83.8±14.5% (Position #1 and three PCs). The reason for this lies in the selection of optimal number of PCs i.e., larger number of PCs yields to higher classification accuracy. Therefore, classification of 6 wrist movements with 5 PCs had higher accuracy than classification of three wrists with three PCs. Another possible reason could be the complexity of three grasping movements. It has been shown that healthy subjects commonly reveal 5 or more synergies in forearm muscles [31]. Consequently, three PC components although satisfied our optimal criteria, still may be not enough to seize the grasping nature.

### 3.2 Visualization of averaged classification accuracy and confusion matrices for re-training QDA

The results of the hand movement classification using optimal number of PCs for QDA classifier re-trained for individual subjects are represented by confusion matrices in Fig. 4, and sensitivity and specificity of the classifier in Table 5.The results are presented separately for three sets of movements and three electrode array positions.

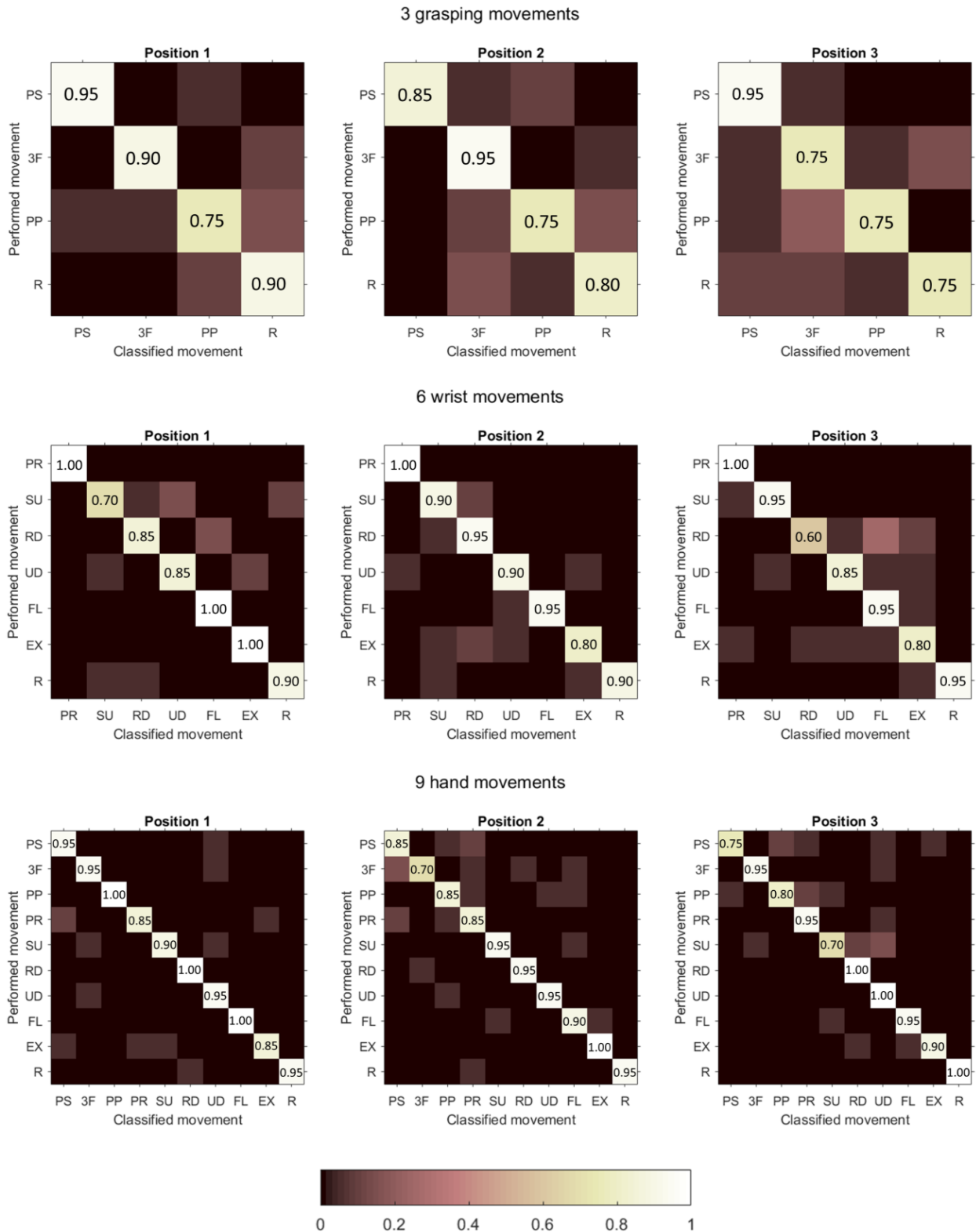

Figure 4, Confusion matrices for grasping (top), wrist (middle), and hand (bottom) movements. Note that "hand movements" refers to all movements. Abbreviations are PS - spherical power grasp, 3F - three finger sphere grasp, PP - two finger prismatic grasp, PR - pronation, SU - supination, RD - radial deviation, UD - ulnar deviation, FL - wrist flexion, EX - wrist extension, and R - resting position. Positions #1-3 signify the array electrode locations. The shade bar for classification accuracy normalized from 0 to 1 is displayed below the confusion matrices.

Table 5, Sensitivity and specificity of QDA classifier re-trained for individual subjects with optimal number of PCs for three positions of electrode array (Position #1-3) and three sets of movements (three grasping, 6 wrist, and 9 hand movements).

|  |  | Position #1 |  | Position #2 |  | Position #3 |  |
|---|---|---|---|---|---|---|---|
|  |  | Sensitivity [%] | Specificity [%] | Sensitivity [%] | Specificity [%] | Sensitivity [%] | Specificity [%] |
| Three grasping movements + Relaxation | PS | 95.0 | 98.3 | 85.0 | 100.0 | 95.0 | 93.3 |
|  | 3F | 90.0 | 98.3 | 95.0 | 90.0 | 75.0 | 88.3 |
|  | PP | 75.0 | 95.0 | 75.0 | 95.0 | 75.0 | 96.7 |
|  | R | 90.0 | 91.7 | 80.0 | 93.3 | 75.0 | 95.0 |
|  | mean±std | 87.5±8.7 | 95.8±3.2 | 83.8±8.5 | 94.6±4.2 | 80.0±10.0 | 93.0±3.6 |
| 6 wrist movements + Relaxation | PR | 100.0 | 100.0 | 100.0 | 99.2 | 100.0 | 98.3 |
|  | SU | 70.0 | 98.3 | 90.0 | 97.5 | 95.0 | 99.2 |
|  | RD | 85.0 | 98.3 | 95.0 | 96.7 | 60.0 | 99.2 |
|  | UD | 85.0 | 97.5 | 90.0 | 98.3 | 85.0 | 98.3 |
|  | FL | 100.0 | 97.5 | 95.0 | 100.0 | 95.0 | 94.2 |
|  | EX | 100.0 | 98.3 | 80.0 | 98.3 | 80.0 | 95.8 |
|  | R | 90.0 | 98.3 | 90.0 | 100.0 | 95.0 | 100.0 |
|  | mean±std | 90.0±11.1 | 98.3±0.8 | 91.4±6.2 | 98.6±1.3 | 87.1±13.8 | 97.9±2.1 |
| 9 hand movements + Relaxation | PS | 95.0 | 97.5 | 85.0 | 95.8 | 75.0 | 99.2 |
|  | 3F | 95.0 | 98.3 | 70.0 | 99.2 | 95.0 | 99.2 |
|  | PP | 100.0 | 100.0 | 85.0 | 97.5 | 80.0 | 98.3 |
|  | PR | 85.0 | 99.2 | 85.0 | 95.8 | 95.0 | 97.5 |
|  | SU | 90.0 | 99.2 | 95.0 | 99.2 | 70.0 | 98.3 |
|  | RD | 100.0 | 99.2 | 95.0 | 99.2 | 100.0 | 97.5 |
|  | UD | 95.0% | 97.5 | 95.0 | 99.2 | 100.0 | 95.0 |
|  | FL | 100.0 | 100.0 | 90.0 | 97.5 | 95.0 | 99.2 |
|  | EX | 85.0 | 99.2 | 100.0 | 99.2 | 90. | 99.2 |
|  | R | 95.0 | 100.0 | 95.0 | 100.0 | 100.0 | 100.0 |
|  | mean±std | 94.0±5.7 | 99.0±1.0 | 89.5±8.6 | 98.3±1.5 | 90.0±11.1 | 98.3±1.4 |

## 4 DISCUSSION

ANN classifier performed slightly better than LDA and QDA for the smallest set of three grasping movements. For 6 wrist movements we did not observe statistically significant difference among classifiers. For the largest set of 9 hand movements, LDA and QDA were distinctly better. When classifier was re-trained for each individual subject, the highest averaged accuracy for all 8 PCs in Position 1 was 99.5% for LDA, 100% for QDA, and 91.9% for ANN. This finding is in line with statement that there is no consensus which classifier performs best for hand movement classification [28, 32].

Leave-one-out cross-validation showed relatively high discrepancies and large degradation when classifier is trained in 9 and tested in one subject. For classification of the largest dataset, the highest classification accuracies (8 PCs) in Position 1 were 67.1%, 66.1%, and 55.6% for LDA, QDA, and ANN, respectively. Although, the obtained accuracies are above chance level (10% for nine movements and relaxation posture), they are far below the acceptable for practical applications. Optimal number of PCs varied vastly for cross-validation procedure: for set of three movements from 3-4 for re-training to 2-7 for cross-validation, for set of 6 movements from 4-5 to 3-8, and for set of all 9 movements from 5-6 to 4-7 PCs (Tables 2-4). This shows that the presented approach is applicable exclusively in cases when classifier is re-trained.

**4.1 Classifier performance in relation to the electrode position for re-training QDA**

Our results showed that the rotation of 2 cm (Position #3) produced somewhat higher dispersion of classification accuracies in relation to the initial Position #1 and compared to the 1 cm translation (Position #2) for re-training QDA. Discrepancies of average accuracy between Positions #1 and #3 were in a relatively small range from -3.8% to 4.5% for three hand movement sets, while absolute differences between Positions #1 and #2 were always positive leading to slightly higher classification accuracies for Position #2 (in a range 0-4%). This somewhat better effect of electrode translation (Position #2) than electrode rotation (Position #3) on classification accuracy might be a consequence of incorporating other forearm muscles more proximal such as *brachio radialis* in relation to our initial Position #1 targeted to the bulks of *extensor carpi radialis* and *flexor carpi ulnaris* muscles. The number of muscles covered by electrode array remains the same when rotation is introduced and might change after translation covering non-identical anatomical landmarks in respect to the underlying musculature (see Fig. 2). Specifically, after rotation, the sEMG array covers the same landmarks with similar crosstalk compared to the translation which has been confirmed in the study that thoroughly assessed sEMG crosstalk in the forearm muscles [33]. Hence, we expected that rotation of the sEMG array electrode would introduce smaller changes in classification accuracy in comparison with translation that did not happen, although for majority of cases those changes were not significant. Two-way ANOVA revealed significant effect of electrode array only in a set of three grasping movements (p = 0.03). For this set, averaged classification accuracy was significantly higher for Position #1 (85.8±17.6%) than for Position #3 (79.5±20.8%). We believe that this result for classification of three grasp movements should be taken with precaution as the relative difference was <8%, the difference of classification accuracies between these two positions for optimal number of PCs showed even smaller relative difference <5%, p value was relatively close the boundary of 0.05 (p = 0.03), and one-way ANOVA for individual sets of features being more important in this study showed no statistically significant differences.

Our method might be applicable to a system involving daily calibration and aiming at accommodation to changes in electrode dislocation as we did not use the data from various electrode array positions to form the training data set. Having in mind main challenges with possible solutions for sEMG-based pattern recognition application in clinics it was suggested that daily calibration can compensate for sEMG electrode impedance, muscle hyper- or hypotrophy, and learning effects [8]. Other successful approaches incorporating sEMG data recorded for electrode shifts in training set for the classifier have been proposed, too. For instance it is reported that when perpendicular and parallel shifts (1 cm and 2 cm) are accounted in the classification feature set during the training phase, the classification error decreases, in some cases reaching substantial error decline of up to ~18 times (from 35.3% to 1.9%) [17]. Another interesting approach studied the effect of the rotation of a circumferentially placed electrode array in clockwise and counterclockwise directions by 1 cm on classification accuracy [34]. Hargrove et al. [34] found that a system trained with features detected from all displacement electrode locations performed better (approx. error 10-20%) than one in which only a single electrode location was used (approx. error 30-40%). In this study, we tested the effect of dislocation on classification accuracy and feature extraction modality, presenting suggested possible scenario from [8].

Average sensitivity and specificity presented in Table 5 show higher specificity than sensitivity in all three electrode array positions and for all movement sets. This indicates that the proposed re-training of QDA is less reliable for distinguishing the proportion of movement samples correctly classified. However, individual results (for specific hand movements) were not consistent. Overall, for the three grasping movements ~40% of sensitivity and specificity parameters were >95%, for 6 wrist movements ~70%, and for 9 hand movements ~80%. This indicates that larger number of optimal PCs preserves discriminatory information even in relatively large dataset (9 hand movements + Relaxation).

**4.2 Classifier performance in relation to the feature set size and selected movements for re-training QDA**

The results from Tables 2-4 for QDA and re-training approach reveal dependency of the classification accuracy on the feature set size, thus confirming our result that the optimal feature set size can be chosen independently of sEMG electrode array position when re-training is implemented and it varies from three to 6. Here, optimal number of features was proportional to the movement set size which is in accordance with the previously published results in [4]. It can be assumed that hand movement set size together with hand movement types within each set influenced the optimal number of PCs.

Different sets of hand movements would naturally result in different misclassifications. In this study, the chosen sets of hand movements led to confusion probably as a result of similar muscle activation for movement execution. Confusion matrices presented in Fig. 4 visualize information with regard to the specific hand movements for the three electrode array positions, three movement sets, when an optimal number of PCs is used for classification. Fig. 4 (top panel) presents a relatively large number of confusions for the three grasping movements (most prominent for Position #3), though the classification accuracies are relatively acceptable >80% (see Table 2). In case of 6 wrist movements (Fig. 4 middle panel), the greatest amount of confusions i.e., the lowest classification accuracy was found for SU (70%) in Position #1, EX (80%) in Position #2, and RD (60%) in Position #3.. PR was the only movement that was successfully detected in 100% of cases for Positions #1-3. In case of the largest set of 9 hand movements (Fig. 4 bottom panel), classification accuracy of different movements varied based on the electrode array position. In Position #1, PR was misclassified as PS or EX in 15% of the cases, while EX was equally misclassified as PS, PR, and SU. Grasp 3F had the lowest classification accuracy (70%) in Position #2, as it was often classified as PS, PR, RD, or FL. The lowest accuracies in Position #3 were in case of PS (75%)and PP (80%) grasps, as well as SU (70%).

Overall, our results indicate that both simple wrist movements and functional grasping movements can lead to both higher and lower confusions. This might be a consequence of muscle anatomy and of the type of chosen hand movement complexity as for example 3F (three finger sphere grasp) incorporates FL (wrist flexion). Previously, SU and PR were proven to be more sensitive to electrode shifts [35]. In our study, PR had relatively higher classification accuracies compared to other hand movements and SU was never confused with FL (in the set of 6 wrist movements) as might be expected. This disagreement is probably the consequence of the classification procedure since our classifier was re-trained in comparison to the classifier in [35]. In addition, PS was classified with higher accuracies and that can be explained by the fact that the sEMG during PS had a different sEMG activation compared to the sEMG for other movements. Recently sEMG-based, kinematic, and general taxonomies for human hand movements revealed, among other results, similarities of sphere grasps, which is confirmed by confusions presented between PS and 3F [36]. Nevertheless, comprehensive comparison with [36] is not possible due to the differences in sEMG measurement setup.

**4.3 Limitations of the presented study**

Although, we proved that classifier re-training can reach acceptable classification accuracy ~90% with application of optimal feature set size and that interaction between electrode array location and the feature set size is not statistically significant, our approach has limitations. In summary, the limitations of the presented study are:

(1) The presented classifier and its robustness in the sense of electrode displacement and optimal set size for sEMG-based pattern recognition have not been checked nor designed for real-time applications. We aimed at exact relation between feature set size and electrode dislocation on classification accuracy that required other measurement conditions to be kept constant in the laboratory setting as their presence could affect the classifier performance [2]. Therefore, we did not check real-time performance, hand posture, load effects, or other known effects [1-2, 10, 37-40]. Consequently, we did not test the effect of windowing on our dataset as crucial step for online implementation [28- 29, 41].

(2) The sEMG data from amputee subjects were not provided for this study, as our research was focused on sEMG-based hand recognition in general, not solely for neuroprosthesis with approach similar to [2]. Previous findings showed no significant differences between healthy subjects and amputees for decoding movements when using sEMG electrodes [1, 42-43] and we believe that the presented results might be used in the field of neuroprosthetics. Nevertheless, precautions should be taken into account in high-level or partial wrist amputees with insufficient remaining musculature, together with other clinical parameters such as phantom limb sensation intensity [3, 10, 44].

(3) We compared only three classifiers (LDA, QDA, and ANN), but we did not perform extensive comparison of classification method as it was performed in [1, 18, 45].

(4) We did not consider sensor fusion. However, it is noteworthy that accelerometry data and sEMG were found to be complementary modalities and significant gains were achieved [39, 46]

(5) We did not check various electrode configurations, as we used bipolar which has been previously recommended [35, 47].

(6) We did not test the effect of feature selection to the classifier performance. Also, we did not check whether application non-linear dimension reduction methods or the exclusion of the PCA would effectively provide dimension reduction or prevent over-fitting [28-29, 48].

## 5 CONCLUSIONS

The main finding of the presented study is that an optimal feature set size and classifier re-training can lead to satisfactory classification accuracy. Therefore, by selecting an optimal number of PCs and with daily re-calibration, the sEMG electrode array can be repositioned (rotated and translated) with no statistically significant effect on hand gesture recognition. Also, we showed that for offline analysis, classifier selection does not play a key role in hand movement recognition, though LDA and QDA performed slightly better than ANN for the largest dataset.

Future work should include enlargement of the current dataset [22] as well as classification results with different positions incorporated in training and testing datasets. Larger dataset would enable further insight into the mismatch between training and testing in relation to the electrode array position [19].

In the future, we plan to test whether relaxation removal increases hand motion discriminability as suggested in [49].

## AUTHOR CONTRIBUTIONS

**N. Miljković:** Conceptualization, Methodology, Data Curation, Writing- Original draft preparation, **M. S. Isaković.**: Conceptualization, Data Curation, Methodology, Investigation, Writing- Review & Editing.

## DECLARATIONS OF INTEREST

None.

## ACKNOWLEDGEMENTS

Funding: The work on this project was partly financed by the Min-istry of Education, Science, and Technological Development, Republic of Serbia (N.M. was partly supported by the grants TR33020 and OS175016, and M.S.I was partly supported by grant OS75016).

Special appreciation the authors owe to Professor Mirjana B. Popović from the University of Belgrade for her kind support, precious guidance, and advice regarding this research which significantly improved the manuscript. Also, the authors would like to thank Dr Matija Štrbac from Tecnalia Serbia Ltd. for providing advice throughout the study. The authors thank all volunteers for their participation.